
%
\magnification=1200
\baselineskip=18pt
\normallineskip=8pt
\overfullrule=0pt
\vsize=23 true cm
\hsize=15 true cm
\font\bigfont=cmr10 scaled\magstep1
\font\ninerm=cmr9
\footline={\hss\tenrm\folio\hss}
\pageno=1
\newcount\fignumber
\fignumber=0
\def\fig#1#2{\advance\fignumber by1
 \midinsert \vskip#1truecm \hsize14truecm
 \baselineskip=15pt \noindent
 {\ninerm {\bf Figure \the\fignumber} #2}\endinsert}

\def\ref#1{$^{[#1]}$}
\def\sqr#1#2{{\vcenter{\vbox{\hrule height.#2pt
   \hbox{\vrule width.#2pt height#1pt \kern#1pt
   \vrule width.#2pt}\hrule height.#2pt}}}}

\noindent{preprint HLRZ ../94}
\bigskip
\bigskip
\centerline{\bigfont SIMULATIONS OF PRESSURE FLUCTUATIONS AND}
\centerline{\bigfont ACOUSTIC EMISSION IN HYDRAULIC FRACTURING}
\bigskip
\bigskip
\bigskip
\bigskip
\smallskip
\centerline{{\bf F. Tzschichholz$^{1,2}$, H.J. Herrmann$^{2,3}$}}
\medskip
\centerline{$^1$ Division of Physics and Mechanics, School of
Technology,}
\centerline{Aristotele University of Thessaloniki, 54006 Thessaloniki,
Greece}

\smallskip
\centerline{$^{2}$ HLRZ, KFA J\" ulich, Postfach 1913,
5170 J\" ulich, Germany}
\smallskip
\centerline{$^{3}$ P.M.M.H. (C.N.R.S., U.R.A. 857), E.S.P.C.I}
\centerline{10 rue Vauquelin, 75231 Paris Cedex 05, France}
\bigskip
\bigskip
\bigskip
\noindent{\bf Abstract} \par
\smallskip
We consider
a two dimensional lattice model to describe the opening of a crack
in hydraulic fracturing. In particular we consider that
the material only breaks under tension and the fluid has no pressure
drop inside the crack. For the case in which the
material is completely homogeneous (no disorder)
we present results for pressure and
elastic energy as a function of time
and compare our findings with some analytic results
from continuum fracture mechanics.
Then we investigate fracture processes in strongly
heterogeneous cohesive environments. We determine the cummulative
probability distribution for breaking events of
a given energetical magnitude (acoustic emission). Further we
estimate the probabilty distribution of emission free
time intervals.
Finally we determine the fractal
dimension(s) of the cracks.

\vskip 2.0truecm\noindent

\leftline{\bf PACS numbers: } 46.30, 91.60.-x, 05.70

\vfill\eject
\noindent{\bf 1. Introduction}
\smallskip

Hydraulic fracturing is widely used
in soil mechanics to improve the permeability of
reservoirs either in oil recovery or of geothermal wells\ref {1}.
An incompressible fluid, in general water, is pushed under high
pressure deep into the ground by
injecting it into a borehole. The fluid penetrates
into the solid opening long cracks.
In order to optimize this rather costly procedure
it is crucial to get some deeper understanding of how
the fracturing occurs.

In the field it is
unfortunately very difficult to obtain direct
information of the crack evolution in the ground. In present
engineering essentially two types of measurements
can be performed: On one hand one can monitor the
pressure fluctuations at the injection pump and on the
other hand one can register acoustic emission signals
on the surface.
In two dimensional Hele-Shaw cells
some controlled laboratory experiments have been
performed~\ref{2} by injecting water
or air into the center of the cell.
The resulting cracks
display a ramified structure which for high enough pressures is
fractal with a dimension of 1.4 - 1.5.

Using a triangular network of springs and radially stretching the network
on the outer boundary into the six directions of a
hexagon the breaking of a material from a central hole
was investigated by several authors~\ref{3}.
They observed fractal cracks having a
dimension that depended very much on the type of applied
displacements (shear, uni-axial, radial).
A stability analysis~\ref{4}
of the boundary of a circular hole with internal pressure
has, however,  shown that this case differs considerably
from that of a stretched membrane due to
the non-linear dependence of the growth velocity of the crack
surface arising from the threshold in cohesion force that must be overcome
to break the material.

We have introduced a model~\ref{5} in which
the imposed load represents a pressure that acts
along the entire (inner) surface of the crack in a direction
perpendicular to the surface (von Neumann
boundary value problem). In this way, the
point of application of the imposed load varies during the
growth of the crack, a situation that
describes the case of hydraulic fracturing
more realistically than previous spring models.
It also turned out to be more efficient to use a beam model
instead of springs.
This model had, however,  two drawbacks: On one hand the pressure was kept
fixed while in real applications it is usually easier to sustain
a fixed injection rate.
On the other hand the heterogeneities of the
medium were ``annealed'', i.e. changing in time while the
disorder in breaking strength or stiffness in real soils is
usually ``quenched'', i.e. constant on the time scales of the
breaking process.

In the present paper we present a model
with constant injection rate in which the variations of pressure can
be measured and in which the cohesion force is a
time independent random variable.
We investigate the strong pressure fluctuations and measure the
energy release as function of the statistical distribution of
cohesion forces.
\bigskip
\noindent{\bf 2. The Model}
\smallskip

In the following we will outline the employed model. First we
give a brief description of the basic elastic equations
and an explanation of how to incorporate heterogeneous
cohesion properties into the fracture model.
After this we explain in detail the used boundary conditions.
Finally we present the employed breaking rules which contain
the physics of the
considered breaking process.

\smallskip

We consider the beam model (as defined in p.~232 of Ref.~6) on
a two dimensional square lattice of linear size $L$. Each of the
lattice sites $i$ carries three real variables:
the two translational displacements $x_i$ and $y_i$ and
a rotational angle $\varphi_i $. Neighbouring sites
are rigidly connected by elastic beams of length $l$.
The beams all have the same cross section  and
the same elastic behavior, governed by three material
dependent constants $a=l/(EA)$, $b=l/(GA)$
and $c=l^3/(EI)$ where
$E$ and $G$ are the Young and shear moduli, $A$ the cross section of
the beam and $I$ the moment of inertia for flexion.
We used for all simulations the values $a=1.0$, $b=0.0017$ and $c=8.6$.
When a site is rotated ($\varphi_i \ne 0$)
the beams bent accordingly always forming
tangentially $90^{\circ}$ angles with each other.
In this way local momenta are taken into account.
For a horizontal beam between sites $i$
and $j$ one has the longitudinal force acting at site $j$:
$F_j = \alpha (x_i -x_j)$;
the shear force:
$S_j = \beta (y_i - y_j) +{\beta\over 2}l
(\varphi_i +\varphi_j)$,
and the flexural torque at site $j$:
$M_j ={\beta\over 2}l(y_i -y_j +l\varphi_j) +
\delta l^2(\varphi_i-\varphi_j)$, using $\alpha = 1/a$,
$\beta=1/(b+c/12)$ and $\delta=\beta(b/c+1/3)$.
The corresponding equations for vertical beams are similar.
In  mechanical equilibrium the sum over all internal and
external forces (torques) acting on site $j$ must vanish
giving rise in the continuum to the Cosserat equations.
We do not consider here inertial or bulk forces, as for example gravity.
\smallskip

Before discussing the employed boundary conditions and
their physical motivation it is convenient to describe how we included
heterogeneous cohesion properties into the fracture model.
The concept of a local cohesion strength has been used
in a number of papers~\ref{7}.
One  assumes that a deformed elastic beam
connecting sites $i$ and $j$ always breaks
above a certain material specific threshold force
$f^{ij}_{coh}$ ('cohesion strength`).
If the applied stresses (forces/beam section) are
above this threshold the beam does break and is removed, i.e. its
elastic moduli are set to zero.
Since the cohesional strength for
compression  is much higher than for tension
we assume~\ref{5}
that compressed beams can {\it never} break.
If all beams have the same cohesion strength the material
is homogeneous.
Such homogeneous
states are usually investigated in continuum fracture
mechanics using the concept of the solid's free
surface energy. In the next section we will present some results
for equal cohesion thresholds.
More realistic, however,  is the case in which the breaking thresholds are
distributed randomly according to some probability density function,
i.e. following a power-law, $\rho (f_{coh})\sim f_{coh}^r$ with
$f_{coh}\in [0,f_{max}]$ and $r > -1$.
Negative exponents $r$ are used to describe strong
cohesive disorder while large positive exponents correspond to weak
disorder.
It is convenient to express
the normalization factor and $f_{max}$ by the distribution's expectation
value $\langle f_{coh} \rangle$ and the exponent $r$.
We fixed
the average cohesion strength for all simulations to be
$\langle f_{coh}\rangle =0.01$ and investigated the fracture processes
for the
exponents: $r =-0.7$, $r=-0.5$ (strong disorder) and
$r=+\infty$ (no disorder).

\smallskip

Boundary conditions must be defined on the external edges of the
lattice and on the internal crack surface.
Concerning the external boundary conditions all
displacements and rotations of the
sites on the outer edges
are assumed to be periodic in horizontal and vertical direction so
that the lattice is spanned on a torus.
Periodic boundary conditions
are preferrable to free boundary conditions
because we are
interested in asymptotic results for large (infinite) systems.
We note that in this case the system cannot expand globally.
The second kind of boundary conditions concerns the conditions for
forces and torques acting on the inner crack surface. While in
tensile experiments the crack surface(s) are always stress free, in
hydrodynamic fracturing these surfaces are loaded by a
pressure distribution resulting from the invading fluid or gas.
In this work we will only consider a homogeneous
(spatially constant) pressure distribution acting perpendicular
along the entire inner crack surface.
However, in contrast to previous simulations\ref{5} here the
pressure has strong fluctuations in time.
Instead of
keeping the inner pressure constant during the simulation we
consider the case where the fluid flux $\Delta V$ into the crack is
constant in time. Hence the crack opening volume $V$ which corresponds
to the total amount of injected incompressibel fluid increases
linear in time $t$,
$$ V(t)=\Delta V  t, \qquad \Delta V = const. ,\eqno{(1)}
$$
a condition which is close to the
situation of industrial hydraulic fracturing.
For completeness we should  mention
that the above mentioned  equivalence
between the  crack opening and the injected fluid volume does only
hold if there is no other sink of fluid in the system besides
the crack.
Although in practice
a loss of fluid in soils is quite common
for seek of simplicity we do not consider here this effect.

Through eq.$(1)$ the crack volume $V$ increases in time. We will
follow the evolution of the crack growth under the continous increase
of loading in the hole. We do this in an iterative way. At each
time step the volume increases by a fixed amount $\Delta V$ and we
estimate from the corresponding elastic solution the stress
distribution on the crack surface. According to the breaking rule
certain beams can break at a given time step depending whether
their stresses are beyond their breaking thresholds or not. When no
beam breaks we continue with the next time step. However, if
beams break they are
simultanous and irreversibly  removed from the set
of elastic equations before one proceeds to the next time step. The
simulation stops when a certain maximum volume $V_{max}$ is reached.
During the simulation we monitor the acting crack pressure, the number
of broken beams and the elastic energy of the system.

In the following we give a more detailed description of the above steps.
Our simulation starts at time $t=1$.
At the place into which the incompressible
fluid is injected (center of the lattice)
one vertical beam connecting sites $i$ and $j$ is removed
from the force and momentum balance equations, i.e. the beam is broken.
Since we want to simulate the loading of a crack by an
injected fluid
a pair of opposite forces (dipole) of unit strength
is applied at the sites $i$ and $j$
pointing into the elastic bulk. The (unit) pressure is then just
defined as the acting force per beam length $l$ ($l \equiv 1$).
Under the influence of the unit
force dipole the lattice becomes distorted.
Lattice distortions are in general characterized
by the displacement field
$\vec{u}(t) =(x_i(t),y_i(t),l\varphi_i(t))$, which in turn is
determined from the boundary conditions at time $t$.
In the following we will denote by $\vec{u}_0(t)$
the displacement field
calculated for a unit pressure $P_0$.
In Ref.~\ref{5} we described how to
calculate the crack
volume $V$ from given crack surface displacements $\vec{u}$.
We denote by $V_0(t)$ the crack volume when
computed from $\vec{u}_0(t)$.
At time $t=1$ the injected fluid
volume $V(1)=\Delta V$ exerts an equilibrium pressure
$P(1)=\Delta V /V_0(1)$ on the crack surface. This follows from
the linearity of the elastic equations.
The elastic solution
which corresponds to this equilibrium pressure is just
$\vec{u}(t) = P(t) \vec{u}_0(t)$. We note that this simple
relation only holds if the acting pressure can be considered as
spatially homogeneous.
\smallskip

We will consider that
only beams along the surface of the inner hole can break.
In that way only one single crack is generated.
At each time step we determine for all
beams on the crack surface the force $f^{ij}(t)$
acting along its axis
and only if this $f^{ij}$(t) is positive, i.e. under tension, and larger
than the breaking threshold $f_{coh}^{ij}$ the beam is broken, i.e.
its elastic
constants $\alpha ,\beta , \delta $ are irreversible set to zero.
The forces $f^{ij}(t)$ must be calculated from the actual
displacements $\vec{u}(t)$.
Note that at a given time step $t$ either several beams can break
simultaneously or none
at all. The latter is the case if the stresses of all
crack surface beams are below their thresholds.
In such a  situation
the crack volume for unit pressure does not change, $V_0(t+1)=V_0(t)$,
because the boundary value problem does not change, i.e.
$\vec{u}_0(t+1)=\vec{u}_0(t)$.
Hence it is not nescessary to recalculate the equilibrium forces.
If a certain number of beams breaks simultanously
additional unit
force dipoles have to be applied at their corresponding neighboring
sites $i$ and $j$ destroying the previous balance of forces.
Then one has to
calculate again the internal equilibrium of
forces. This is done
in our case using a conjugate gradient algorithm with a precision of
$\varepsilon =10^{-10}$ (see Eq.~(47) in Ref.8).
Due to the new boundary conditions
the elastic solution changes, i.e. $\vec{u}_0(t+1)\ne\vec{u}_0(t)$,
and the volume $V_0(t+1) \ne V_0(t)$ is recalculated from
the new crack surface
displacements $\vec{u}_0(t+1)$, as described in ref.~\ref{5}.

At time $t+1$ the crack pressure takes the value $P(t+1)=V(t+1)/V_0(t+1)$
and the elastic solution matching eq.$(1)$ is given by
$\vec{u}(t+1) = P(t+1) \vec{u}_0(t+1)$.
{}From these displacements one recalculates the forces $f^{ij}(t+1)$
for all beams on the (new) crack surface, decides which are
the next beams to be broken and repeats the above described procedure.

\bigskip
\noindent{\bf 3. Fracture of homogeneous solids}

\smallskip

Although homogeneous cohesive properties in solid systems
are rather the exception than the rule it is useful to study
them mainly because of their relative simplicity. Our model should
be capable to reproduce some general features of simple continuum models
for hydraulic fracturing. First we consider the limiting case of
equal breaking thresholds $\langle f_{coh}\rangle = f_{coh}^{ij}
=0.01$ for a lattice of linear size $L=150$.
The crack volume (total amount of injected fluid)
is increased at a
constant rate of $\Delta V=0.05 l^2$ per time step. The
simulations stop when the lattices are
broken into two pieces. As expected we find  the cracks to have a  linear
shape. This is due to the fact that the highest stress
enhancements occur at the two vertical beams at the tips.
Fig.~1 shows the time dependence of the pressure $P$
inside the crack in a
double logarithmic plot. One can see that on average
the pressure drops in time
and has oscillations on short time scales. At the beginning,
i.e. for time $t=1$,
the crack is very small (one vertical broken beam)
and one needs a high pressure
in order to push the fluid of volume $\Delta V$ into the crack.
In this particular calculation the two vertical beams on the crack
surface are already stressed beyond their cohesion threshold.
At the next time step, $t=2$, these two beams are broken and the
pressure drops, because the grown crack can now be
opened much easier than before (see fig.~1). The pressure goes down
although additional fluid $\Delta V$ has been added to
the crack at this time step.
It is obvious that a large crack experiences
a much lower pressure than a small crack for the same opening  volume
because the system in that case is locally more stressed.
If the pressure drops too much (like at time $t=2$) the
stresses at the two crack tips fall below their cohesion value and the
crack cannot grow at the next time step.
By injecting more fluid into the crack the pressure
increases linearly in time until the cohesion forces can be overcome
again. In fig.~1 one  sees a sequence of this oscillating
pressure.
In the continuum description a smooth decrease
of the breaking pressure
in time (volume) is expected.
Using continuum mechanics one can argue that the smallest
pressure necessary to extend the crack at a given opening volume
should behave like $P_{crit}\sim V^{-1/3}$ in $d=2$ \ref{9}.
Such a relationship should only hold for an elastic {\it infinite}
plate. Because
of eq.$(1)$ we can identify crack volume with time up to the
factor $\Delta V$.
For comparison we have plotted in fig.~1 a dotted line having a
slope of $-1/3$. The agreement with our numerical values is quite
acceptable over one decade. For small and large times we obtain
pronounced deviations which originate from the
lattice structure and the finite size of the lattice
and from the external boundary conditions.

It is interesting to consider the temporal evolution of the stored elastic
energy $U$. We have calculated the time dependent energy
directly by summing
up the elastic energies of all non broken beams in the
system at every time step. In fig.~2 we show in a log-log plot the
time dependence of the elastic energy.
Again we see an oscillating behavior as
discussed above for the pressure. Breaking
events as for example at time $t=1$ and $t=3$ {\it decrease} the
elastic energy while it is {\it increased}
by pushing fluid into the crack.
Globally the elastic energy must increase because the system
becomes more compressed in time.
We find that the peak energies
scale in time as a power law
over two and a half decades with an
exponent close to $2/3$, see fig.~2. This exponent can be understood
by calculating the work
$W_{crit}= \int_0^V P_{crit}(V')\quad dV'$ done by the
external forces in order to inject the fluid volume $V$, which yields
$2U_{crit}= W_{crit}\propto V^{2/3}$.
This agrees well with our numerical findings.
Again finite size effects lead to
deviations from the behavior of an infinite continuum.

\bigskip
\noindent{\bf 4. Cracks in heterogeneous solids}

\smallskip

In the following we will consider
the influence of  strongly heterogeneous cohesive strengths on
the hydraulic fracturing process. It is well known that
fracture processes in disordered solid systems show a rich
phenomenology concerning the crack geometry as for example
crack branching and crack deflection\ref{6}.
There exist phenomena, as for example the appearence of
microcracks in solids,
which cannot be explained by equilibrium thermodynamics
without considering the heterogenity of
physical properties.
Experimental and theoretical investigations
during the last decades have manifested that the overwhelming number
of fracture phenomena in solid systems is strongly influenced
or even controlled by inhomogeinities.

We will investigate the statistical properties
of breaking sequences during hydraulic fracturing and their
correlations in space, time and magnitude.
Similar quantities are frequently  considered in the analysis of
earthquake occurrences \ref{10, 11} and of acoustic emission (AE) during
microfracturing of rocks \ref{12} or of technical materials \ref{13}.
Acoustic emission records for the hydraulic fracturing in geothermal
wells have been published for example in Ref.~\ref{14}.

\smallskip

In our simulations
we have considered threshold distributions for two different exponents
$r=-0.7$ and $r=-0.5$
(see the previous section for the definition of the distribution).
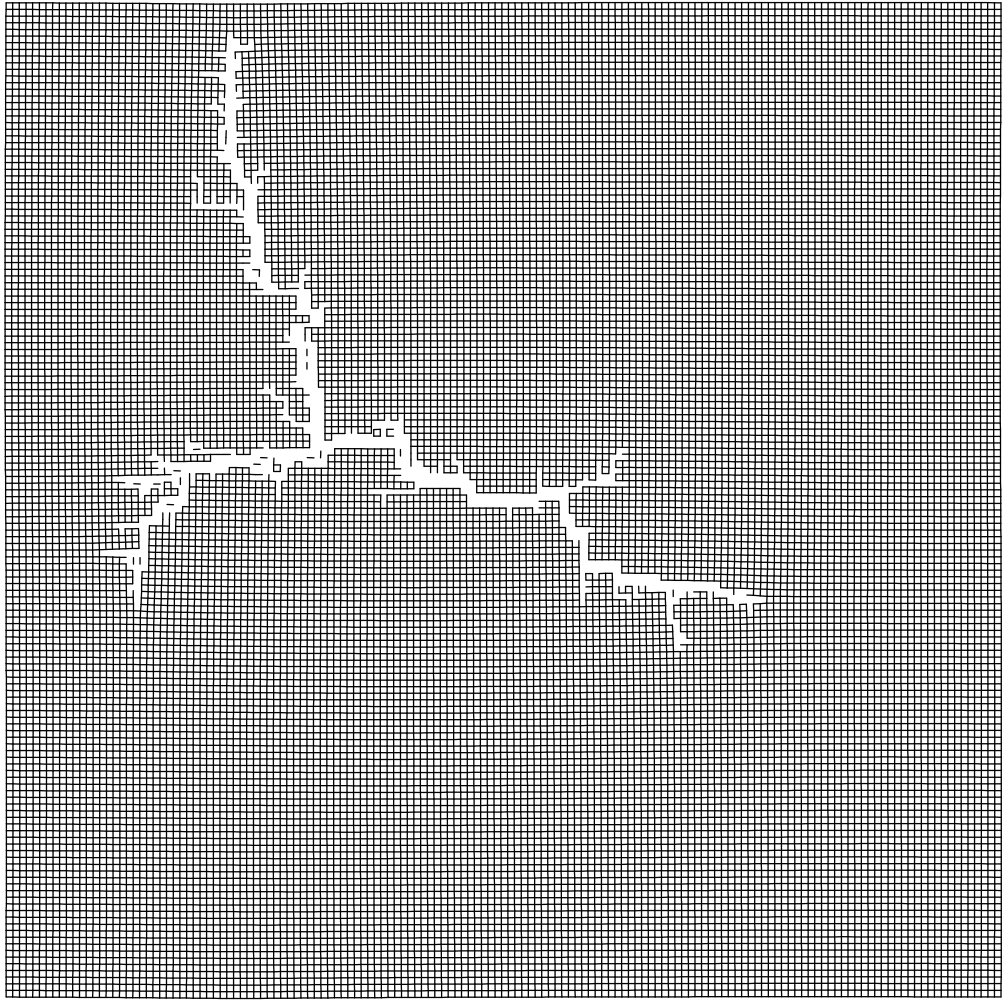
Fig.~3 shows a typical hydraulic crack pattern grown
in a medium with strong disorder ($r=-0.7$)
on a lattice
of linear size $L=150$. The crack consists of $629$
broken beams after $1500$ time steps.
Apparently smaller crack branches appear along larger branches,
a geometrical property
which is typically observed for self similar (fractal) structures.
We have evaluated the fractal dimension(s) of the
generated hydraulic cracks
by considering the relationship  between the typical crack radius
$R(N)$ and the number of broken beams $N$.
For this we  calculate the squared radius of gyration
$R^2(N)={1\over N}\sum_i (\vec{r}_i - \vec{r}_0
)^2$ with $\vec{r}_0 = {1\over N}\sum_i \vec{r}_i$.
Finally the radius $R(N)$ is averaged over
a number of independent configurations in order to obtain results for
{\it typical} cracks. In fig.~4 we show in a log-log plot the
number of broken beams $N$ versus the typical
crack radius $R(N)$ for the two statistics of the heterogenities
characterized by exponents $r=-0.7$ and
$r=-0.5$. We
find in  both cases power laws $N\propto R^{d_f}$ with $d_f > 1$,
giving evidence for fractal crack growth. The exponent $d_f$ is
called the fractal dimension of the crack and takes for
$r=-0.7$ $(\diamond)$ and $r=-0.5$ $(+)$ the values
$d_f=1.44\pm 0.10  $ and $d_f=1.39\pm 0.10 $ respectively.
These values for $d_f$
are consistent with the fractal dimension found in
the two dimensional experiments on hydraulic fracturing of viscoelastic
clays \ref{2}. One detects, however,  from both curves in fig.~4
a crossover at large $R$
to a lower slope $d_f \approx 1.25$. It is interesting
to note that the crossover  radii $R_\times$
depend on the width of the threshold distribution. The
cracks for stronger disorder $r=-0.7$
show a higher crossover radius $R_\times \approx 18$ than those
for $r=-0.5$ with $R_\times \approx 10$.
This is qualitatively in agreement with the observation that broader
threshold distributions have larger homogenization volumes\ref{15}.
In our case it is, however, likely that the crossover behaviour
is an effect of the finite size of the lattice.
It has been argued \ref{15,16} that for logarithmically
diverging threshold distributions, i.e. for $r=-1$, a  fracture
process essentially reduces to a percolation problem. We note
that this is not necessarily the case here
because the employed breaking law
is  {\it asymmetric} with respect to tension and compression.

\bigskip
\noindent{\bf 5. Bursts and temporal correlations}

\smallskip

Since the fractal nature is a finger print for infinite-range
correlations in the crack geometry, one can also ask how
the breaking events are correlated in {\it time}. To illustrate
this question we show in fig.~5 the complete breaking sequence
of the crack displayed in fig.~3. We have plotted the number
of beams broken between two consecutive time steps as a function
of time. Most striking is the fact that the breaking process is
very discontinuous in time. There are large time intervals in which
no breaking occurs at all. During such time intervals of {\it quiescence} all
beams on the crack surface are stressed below their cohesion thresholds
and the acting pressure increases linearly in time. The second
striking fact is that if a breaking event happens
after a period of quiescence it usually triggers a sequence
of consecutive breaking events (temporal clustering).
We will call such sequences {\it bursts}. The bursts themselves are,
as one can see in fig.~5, unequally distributed in time.
They occur relatively often for small times and become more rare later.
However, while the bursts are narrow for early time steps
they typically become broader with increasing time.
The numbers of simultanously broken beams per time step
also exhibit particularities. Let us call these numbers
the {\it magnitudes}
of the breaking events. Broad bursts typically consist of few
events of high magnitude and much more of low magnitude.
Fig.~6 shows the magnifications of two bursts from fig.~5. The ordinate
gives  the number of simultanously broken beams during a time step
and the corresponding 'released energies` as well. The definition of
the released energies is given further below in Sec.~6. We see that
the peak energy releases do {\it not} temporarly coincidence with the peak
numbers of broken beams. In fig.~7 we show the time dependent pressure
belonging to the breaking process shown in fig.~5 and fig.~3 respectively.

The reader familar with magnitude records of earthquakes or of
accoustic emission records from laboratory experiments will
recognize some ressemblance with our data.

In the following we will investigate more closely the temporal
clustering of breaking events.
We have calculated the lifetime $\tau$ for each burst
as the time that elapses between the first and the last breaking
event of a burst.
Fig.~8 shows the (unnormalized) histogram
of {\it burst lifetime}
in a double logarithmic plot. Small bursts occur relatively often
while larger bursts are less frequent.
The statistics are made over $729$ bursts from $60$ samples
for $r=-0.7$ and over $862$ bursts from $53$ crack simulations for
$r=-0.5$. The largest detected burst has a lifetime of
approximately $\tau=140$. We note that for large lifetimes,
$\tau\ge 30$, the statistic becomes unreliable because the occurrences,
$n(\tau)$, become too  small, $n(\tau) < 10$. This is also due to
the fact that all simulations were stopped after $t_{max}=1500$
time steps.
In order to extract more information from the lifetime
distribution we consider in fig.~9 the less noisy cummulative
probability distribution $p(\tau)$ that a given burst has a lifetime
shorter or equal to $\tau$.
With an intermediate range
we observe in fig.~9 that the cummulative probability
$p(\tau) =\sum_{i=1}^\tau n(i) / \sum_{i=1}^{t_{max}} n(i)$
seems to obey a power
law, $p(\tau)\propto \tau^{1-\eta}$, with $\eta =0.54\pm 0.15$.
Hence the lifetime distribution of bursts also follows a power law
in this regime,
$$ n(\tau) \propto \tau^{-\eta}.
\eqno{(2)}
$$
It is remarkable that the two curves in fig.~9 give nearly the
same exponent $\eta$ although the exponents $r$
from their threshold distribution are different. This indicates
that the underlaying generating mechanism for  the temporal
clustering of breaking events might be universal.
In fig.~9  one also sees strong deviations from the power
law for small ($\tau < 4$) and for large ($\tau >30$) bursts
particularly in the
case of $r=-0.7$. The existence of an upper cut off is plausible,
because we stopped each crack growth after $t_{max}=1500$ time
steps. This
artificially lowers the number of large bursts.
The lower cut off is quite common in cluster statistics, known as
'corrections to scaling`.

Next we consider the lifetime
distribution $q(\tau)$ of {\it quiet intervals}. This distribution
characterizes the arrangement of bursts on the natural time scale $t$
and has attracted  interest in seismology because
of its role for possible predictions of earthquakes\ref{17}.
It has been observed that if certain regions undergo high magnitude
earthquakes they often show thereafter rather long periods of seismic
quiescence (inactivity).
We have investigated this statistics
from our hydraulic fracturing calculations.
In fig.~10 we consider in a semi-log plot
the cumulative probability $Q(\tau)$ that two
{\it consecutive} bursts are separated by a
time interval of quiescence shorter or equal equal than $\tau$.
More precisely, we define
the width of the interval $\tau$ by the number of breaking free
time steps between two consecutive bursts.
For intermediate time scales
the probability follows, as shown
in fig.~10, a logarithmic dependence, $Q(\tau) \propto \ln \tau$.
Thus  the probability $q(\tau)$ to find a quiet
time interval of width $\tau$ between two adjacent bursts scales
in this regime as,
$$ q(\tau) \propto {1\over \tau}.
\eqno{(3)}
$$
Interestingly both distributions $n(\tau)$ and $q(\tau)$
scale for intermediate times as power laws, however, with
different exponents.

So far we have considered the lifetime statistics of bursts and
of quiet intervals.
Valuable information about the temporal evolution of hydraulic
cracks can be gained by considering the time
correlations between the breaking events.
{\it A priori} it is not obvious which of the breaking events are
causually connected.
In our model the elementary consecutive breaking events define the
shortest accessible time scale. From acoustic emission
experiments and seismic records, however,  the existence of
a background noise is well established. Events are usually only
considered if their magnitude exceeds the background noise by orders
of magnitude \ref{13, 18}.
The cut off of the background noise leads to
discrete magnitude-time sequences.
This is similar in our model because the discrete nature of the beams
put a lower cut off on the possible size of an event.
Experimentally, the variations in emission magnitudes are so strong
and the line-widths are so small that one can consider
acoustic emission 'events` as delta peaks. We note that
in experiments the
significant breaking events are already temporal clusters (bursts).
Hence one can investigate the time correlations on the scale of a
typical line-width (burst lifetimes)
or on a larger time scale.
First we will investigate the case in which
only correlations {\it within} the bursts are considered.
We have calculated from all bursts
the probability distribution $b(\tau)$ of
time intervals $\tau=\vert t_i-t_j \vert$ between all possible
pairs of breaking events belonging to  the {\it same} burst.
We show in fig.~11 in a semi-logarithmic plot the distribution of
time intervals $\tau$. One clearly sees an exponentially decreasing
probability (two-point correlation) in time, $b(\tau)\propto
\exp{(-a\tau)}$. In this plot one does not find any power-law behavior
for small times, $\tau \le 30$, as one might expect from the power-law
scaling of burst lifetimes, see eq.$(2)$.
In our case this is due to the fact
that the 'large` bursts ($\tau >30$) having a bad statistics
completely dominate the correlations also for small $\tau$.
This appears to be an additional complication resulting from the
{\it very broad} lifetime distribution $n(\tau)$ of bursts,
i.e. the small value
for the exponent $\eta \approx 0.5$.
We can, however, as discussed above
consider the correlations between all breaking events
regardless of the bursts they belong to.
This is conveniently done \ref{10} by calculating the histogram of time
intervals $\tau$ between {\it all possible pairs}
of breaking events $\tau=\vert t_i-t_j \vert$
from a given time record.
In fig.~12 we show in a double logarithmic representation
the normalized histogram of these time intervals.
We find that the probability $g(\tau)$  to detect
two breaking events seperated by a
time interval $\tau$ decreases
on intermediate scales ($15< \tau < 130$ for
$r=-0.7$) as a power law,
$$ g(\tau) \propto \tau^{-\kappa}, \quad \kappa = 0.94\pm 0.20 .
\eqno{(4)}
$$

This corresponds to Omori's well known  law of aftershocks
which has been first formulated in 1894 for  earthquakes \ref{19}.
It says that the probability for an aftershock
following a large earthquake decays as $1/\tau^\kappa$ in time
with an exponent $\kappa$ close to one.
Omori's law
has been verified from earthquake catalogs for
aftershocks series ranging from a few hours
to a couple of years after the main event. The  empirical value
for the exponent $\kappa$ lies between $1.0$ and $1.4$ \ref{10}.
Though the appearance of earthquakes and aftershock series
underlies quite different constraints concerning the rheology and the crack
opening mode than in our two dimensional model we
find a value for $\kappa$ very close to the empirical one in three
dimensions.
Recently it has been argued \ref{18} that the value $\kappa =1$
should hold quite generally for nonlinear fracture problems, such as
earthquakes, independent on the spatial dimension \ref{20}.
Hence one might expect the basic mechanism for bursts sequences,
respectively aftershocks, to be {\it universal} due to {\it self
organisation} of rupture. Physically one might argue that a main
breaking event with high magnitude triggers smaller magnitude events
which in turn create even smaller events {\it ad infinitum}
if the cohesion properties are sufficient
heterogeneous. However, in our case
during a single burst the solid undergoes a quite complicated
stress redistribution process until all microstructural elements
(beams) are below their cohesion threshold. When the crack
grows into new regions
it may be stopped due to a region with
particular high breaking thresholds. Then the burst terminates.
As the crack becomes longer  its critical opening volume
necessary to overcome the thresholds increases even more because
larger cracks can be opened easier than small cracks. This explains
why the intervals of quiescence become longer for larger cracks.

\vfill\eject
\bigskip
\noindent{\bf 6. Released energies, acoustic emission}

\smallskip


So far we have considered the spatial and temporal correlations
of breaking events
during hydraulic fracturing. We have found qualitative and quantitative
similarities to the seismic clustering of earthquakes and to
acoustic emission laboratory experiments.
However, we have so far not discussed
the correlations of the magnitudes (intensities) of
breaking events. For this the
magnitude occurrence relationship is of
central importance \ref{21}.
It has been found empirically that the cumulative
occurrence $H(m)$ of earthquakes of magnitude
larger than $m$ follows
the celebrated Gutenberg-Richter law, $H(m)\propto 10^{-bm}$ \ref{22}.
The 'b-value` found in Gutenberg-Richter's law is often close
to one \ref{10}. The magnitudes of seismic events are defined
by the logarithmic wave amplitudes. The energy release, $\delta U$,
is usually considered to be proportional to the squared
amplitude of the earthquake. One typically obtains a cumulative
occurrence-energy relation of the form, $H(\delta U)\propto {\delta
U}^{-0.6}$, expressing the presence of earthquakes on {\it all}
energetical scales \ref{11, 40}.
It has been shown recently \ref{13}
that acoustic emission (AE) in the ultrasonic frequency range
due to microfracturing of synthetic plaster also exhibits a power law
for the cumulative occurrences with an exponent close to the one
found in the Gutenberg-Richter law.
Similar findings from AE records have been published
earlier for the microfracturing of rocks \ref{23}.

In the following we will discuss some energetical aspects of our model.
The thermodynamical description of crack spreading phenomena
in heterogeneous environments becomes
quite complicated and we will calculate the involved energies
exclusively from
the mechanical equilibrium conditions and {\it not} from thermodynamical
considerations.
In order to determine in our model the amount of energy released
by a given breaking event one has to calculate the stored
elastic energy $U(t)$ at each time step. This energy is, as
already mentioned, determined prior to breaking from the sum over the
energies of all non broken beams. We define the rate of
change of elastic energy as $\delta U=-\partial U/ \partial t =
-(U(t+1)-U(t))$.
If the rate $ \delta U $ is negative
then the lattice has increased its elastic energy.
We have verified that the overwhelming number of breaking events are
detectable by the condition $\delta U > 0$. There are very few events
that show a negative energy flux because the value of $\delta U$
describes the sum of two effects. First, it describes the lowering of
elastic energy due to a breaking event. Second, it describes
the increase of elastic energy due to the added amount $\Delta
V$ of fluid per time step. However, if the cracks and their corresponding
opening volumes become large compared to $\Delta V$ the fraction of
non-detected events becomes very small.
Experimentally the energy dissipation due to crack growth has at least
two contributions. On one hand the creation of additional crack surface
consumes energy.
On the other hand elastic waves are emitted from
the crack tip(s) (acoustic emission (AE)).
Because we do not consider the full dynamic equations it is
a priori not clear which fraction of the dissipated energy to
assign to the 'acoustic emission`. We propose that, as a first
approximation, the AE energy per event is proportional to the
rate of change of elastic energy $\delta U$ of the system.
We show in fig.~13 in a semi-log plot the cumulative occurrence
$H(\delta U)$ for breaking events of energetical magnitude larger than
$\delta U$. We find an exponential relationship,
$$H(\delta) \propto \exp{(-\beta \delta U)}.\eqno{(5)}
$$
The flat tail for large $\delta U$ is due to low statistics. This
result is at variance with what is expected from the Gutenberg-Richter
law and AE measurements in laboratory experiments.
A possible source for this discrepancy could be our definition of
the released energy: We proposed that the AE be proportional
to the total energy relaxation. It could, however, be that the
fraction of energy emitted acoustically is a more complicated function
of the energy. Other sources for the deviations between our model and
experiments could be the periodic boundary condition and the two
dimensionality of the lattice used in our simulations.

\bigskip
\noindent{\bf 7. Conclusions}

\smallskip

We have presented a lattice model for hydraulic fracture which takes
into account the particular boundary conditions inside the hole and
the quenched nature of the heterogeneities in the rock.
We drive the process by increasing the crack volume linearly in time
(constant flux). The pressure fluctuates erratically similarly
to what is measured at boreholes in the field.
The crack shapes are fractal and the fractal dimension agrees well
with measurements performed in Orl\`eans with Hele Shaw cells.
The sequence of breaking events is organized in bursts which have
a life time distribution and a distribution of quiescence times
that are power laws. This indicates that self-organized
criticality (SOC) \ref{24} takes place.
Events inside a burst seem uncorrelated while long range
correlations between bursts follow Omori's law for aftershocks.
The distribution of released energies does not follow a power-law
within the numerical accuracy of our calculations, shading doubts on
the simple hypothesis that acoustic emission signals are simply
proportional to the released potential energy.

This work is still rather preliminary if the full reality of
hydraulic fracturing in oil or geothermal reservoirs is to be
described. Real soils are three dimensional and the distribution
of heterogeneities usually follows a Weibull distribution
with a material dependent exponent. The pressure of the fluid in the
crack is not constant but depends on the distance from the
injection site and the geometry of the crack. The system is nearly
infinite in size and the restrictions in the total volume of the
system as imposed by the periodic boundary conditions are not
realistic. Still many features have been found in this paper
that agree with experimental measurements and we think that
including more details into the model, which is rather
straightforward, could help simulate hydraulic fracturing rather
accurately in the future.

\vfill\eject
\bigskip
\noindent{\bf References}
\medskip
\item{1.} H. Takahashi and H. Ab\'e, in {\it Fracture Mechanics
of Rock}, ed. B.K. Atkinson (Academic Press, London, 1987), p. 241
\item{2.} H. Van Damme, E. Alsac and C. Laroe, C.R. Acad. Sci.
S\' erie II {\bf 309} (1989) 11; H. Van Damme,
in: {\it The Fractal Approach to Heterogeneous Chemistry},
ed. D. Avnir (Wiley, New York, 1989) p.199; E. Lemaire, P. Levitz,
G. Daccord, and H. Van Damme, Phys. Rev. Lett. {\bf 67} (1991) 2009
\item{3.} E. Louis and F. Guinea, Europhys. Lett. {\bf 3} (1987) 871;
E. Louis, F. Guinea and F. Flores, in: {\it Fractals in Physics},
eds. L. Pietronero and E. Tossatti (Elsevier, Amsterdam, 1986);
P. Meakin, G. Li, L.M. Sander, E. Louis and F. Guinea,
J. Phys. A {\bf 22} (1989) 1393;
E.L. Hinrichsen, A. Hansen and S. Roux, Europhys. Lett.
{\bf 8} (1988) 747
\item{4.} H.J. Herrmann and J. Kert\' esz, Physica A {\bf 178}
(1991) 227; H.J. Herrmann, in: {\it Proceedings of the NATO workshop
``Growth Patterns in Physical Sciences and Biology''} (1992)
\item{5.} F. Tzschichholz, H.J. Herrmann, H.E. Roman and
M. Pfuff, Phys. Rev. B {\bf 49}, 7056, (1994)
\item{6.} H.J. Herrmann and S. Roux (eds.) {\it Statistical Models
for the Fracture of Disordered Media} (North Holland, Amsterdam,
1990)
\item{7.} M. Sahimi and J.D. Goddard, Phys. Rev. B {\bf 33}, 7848,
(1986).
F. Family, Y.C. Zhang and T. Vicsek, J. Phys. A {\bf 19}, L733 (1986).
B. Kahng, G.G. Batrouni, S. Redner, L. de Arcangelis and
H.J. Herrmann, Phys. Rev. B {\bf 37}, 7625 (1988).
S. Roux, A. Hansen, H.J. Herrmann and E. Guyon, J. Stat. Phys. {\bf
52}, 251 (1988). L. de Arcangelis, A. Hansen, H.J. Herrmann and
S. Roux, Phys. Rev. B {\bf 40}, 877 (1989).
A. Hansen, S. Roux and H.J. Herrmann, J. Physique {\bf 50}, 733
(1989).
H.J. Herrmann, A. Hansen and S. Roux, Phys. Rev. B {\bf 39}, 637
(1989).
L. de Arcangelis and H.J. Herrmann, Phys. Rev. B {\bf 39}, 2678
(1989).
F. Tzschichholz, Phys. Rev. B {\bf 45}, 12691 (1992).
E. Schlangen and J.G.M. Van Mier, Materials and Structures {\bf 25},
534 (1992).
\item{8.} G. G. Batrouni and A. Hansen, J. Stat. Phys.
{\bf 52}, 747 (1988)
\item{9.} From continuum mechanics  one obtains the relation between
the crack volume $V$, the acting pressure $P$
and the crack length $2R$, $V = {2E\over\pi} PR^2$ in $d=2$.
One the other hand Griffith's
criterion for brittle fracture predicts a relation for the
equilibrium pressure of the form
$P_{crit}= \sqrt{{2E\gamma_0\over{\pi R}}}$, where $E$ and $\gamma_0$
denote the Young modulus and the specific surface energy,
respectively. [See for example: I. Sneddon and M. Lowengrub,
{\it Crack problems in the classical theory of elasticity}.
(SIAM, Philadelphia, (1969))].
By eliminating the crack length from above relations one
obtains the mentioned relation $P_{crit}\sim V^{-1/3}$ for an infinite
system.
\item{10.} See for example: L. Knopoff in {\it Disorder and Fracture},
eds. J. Charmet et al. (Plenum Press, New York, 1990), p. 279-87 and
references therein.
\item{11.} G. Korvin in {\it Fractal Models in the Earth Sciences}
(Elsevier,
Amsterdam - London - New York - Tokyo, 1992) p. 156-67 and references
therein. D.L. Turcotte, {\it Fractals and chaos in geology and
geophysics} (Cambridge Univ. Press, U.K., 1992).
\item{12.} I.G. Main, P.R. Sammonds and P.G. Meredith, Geophys. J.
Int. {\bf 115}, 367-80 (1993). C.G. Hatton, I.G. Main and P.G.
Meredith, J. Structural Geology {\bf 15}, 1485 (1993).

\item{13.} A. Petri, G. Paparo, A. Vespignani, A. Alippi and
M. Costantini, (1994), preprint.
\item{14.} H. Takahashi, H. Niitsuma, K. Tamakawa, H. Ab\'e, R. Sato
and M. Suzuki, Proc. 5th Int. AE Symp. Tokyo, pp. 443-53, (1980).
J. N. Albright and C. F. Pearson, Soc. Petrol. Eng. J., 523, (1982).
H. Niitsuma, K. Nakatsuka, N. Chubachi, H. Yokoyama and
M. Takanohashi, Geothermics {\bf 14}, 525, (1985). H. Niitsuma,
Int. J. Rock Mech. Min. Sci. \& Geomech. Abstr. {\bf 26}, 169, (1989).
\item{15.} A. Hansen in Ref.~6, p. 115-58.
\item{16.} S. Roux, A. Hansen, H.J. Herrmann and E. Guyon,
J. Stat. Phys. {\bf 52}, 237, (1988).
\item{17.} T. Yamashita and L. Knopoff, Science (1989).
S.L. Pepke, J.M. Carlson and B.E. Shaw, J. Geophys. Res.
{\bf 99}, 6769 (1994).
\item{18.} B.E. Shaw, Geophys. Res. Lett. {\bf 20}, 907 (1993).
\item{19.} F. Omori, J. Coll. Sci. Imper. Univ. Tokyo, {\bf 7}, 111,
(1894).
\item{20.} Though the independence of $\kappa =1$ from the spatial
 dimension $d$ has not been stated explicitly in Ref.~18 it follows
from the fact that the given
physical arguments do not depend on $d$.
\item{21.} J. Lomnitz-Adler, J. Geophys. Res. {\bf 98}, 17745 (1993).
\item{22.} G. Gutenberg and C.F. Richter, Ann. Geophys. {\bf 9}, 1 (1956).
C.F. Richter, {\it Elementary Seismology} (W.H. Freeman and
Co, San Francisco, 1958).
\item{23.} C.H. Scholz, Bull. Seismol. Soc. Am. {\bf 58}, 1117 (1968).
C.J. All\`egre, J.L. Le Mouel and A. Provost, Nature {\bf 297}, 47
(1982). C.H. Scholz, Nature {\bf 338}, 459 (1989).
\item{24.} P. Bak, C. Tang and K. Wiesenfeld, Phys. Rev. Lett. {\bf 59}
(1987) 381

\vfill\eject

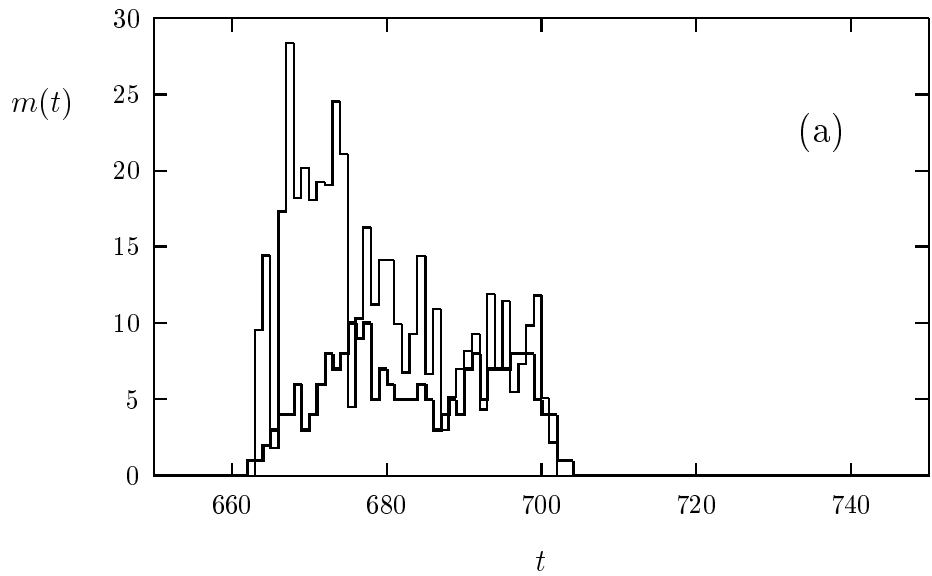
\end